\documentclass[12pt]{article}
\usepackage{graphicx}
\usepackage{amsmath}
\usepackage[utf8]{inputenc}

\newcommand\pubnumber{NuPhys2017-Migenda}
\newcommand\pubdate{\today}

\def\Title#1{\begin{center} {\Large #1 } \end{center}}
\def\Author#1{\begin{center}{ \sc #1} \end{center}}
\def\Address#1{\begin{center}{ \it #1} \end{center}}

\newcommand\pubblock{\rightline{\begin{tabular}{l} \pubnumber\\
         \pubdate  \end{tabular}}}
\newenvironment{Abstract}{\begin{quotation}  }{\end{quotation}}
\newenvironment{Presented}{\begin{quotation} \begin{center} 
             PRESENTED AT\end{center}\bigskip 
      \begin{center}\begin{large}}{\end{large}\end{center} \end{quotation}}

\def\beq{\begin{equation}}
\def\eeq#1{\label{#1}\end{equation}}
\def\eeqn{\end{equation}}

\def\beqa{\begin{eqnarray}}
\def\eeqa#1{\label{#1}\end{eqnarray}}
\def\eeqan{\end{eqnarray}}

\let\bar=\overbar

\def\Dslash{\not{\hbox{\kern-4pt $D$}}}
\def\dslash{\not{\hbox{\kern-2pt $\del$}}}

\def\msb{{\bar{\ssstyle M \kern -1pt S}}}

\begin{document}
\begin{titlepage}
\pubblock

\vfill
\Title{Supernova Burst Observations with DUNE}
\vfill
\Author{Jost Migenda for the DUNE collaboration}
\Address{University of Sheffield, Department of Physics \& Astronomy, Sheffield S3~7RH, UK}
\vfill
\begin{Abstract}
The Deep Underground Neutrino Experiment (DUNE) is a 40-kton under\-ground liquid argon time-projection-chamber detector that will have unique sensitivity to the electron flavor component of a core-collapse super\-nova neutrino burst. We present expected capabilities of DUNE for measurements of neutrinos in the few-tens-of-MeV range relevant for super\-nova detection and the corresponding sensitivities to neutrino physics and supernova astrophysics. Recent progress and some outstanding issues will be highlighted.
\end{Abstract}
\vfill
\begin{Presented}
NuPhys2017, Prospects in Neutrino Physics\\
Barbican Centre, London, UK,  December 20--22, 2017
\end{Presented}
\vfill
\end{titlepage}
\def\thefootnote{\fnsymbol{footnote}}
\setcounter{footnote}{0}

\section{Introduction}
The Deep Underground Neutrino Experiment (DUNE,~\cite{CDR}) is a next-generation liquid argon time-projection-chamber (LAr TPC) detector, whose broad physics programme covers many areas of particle and astroparticle physics. It will improve upon limits from previous nucleon decay searches in various channels and act as a far detector for a neutrino beam coming from Fermilab at a distance of 1300\,km, enabling precision measurements of neutrino oscillation parameters and searches for CP violation in the neutrino sector. DUNE will also study neutrinos from natural sources such as atmospheric neutrinos and supernova neutrinos.

In section~\ref{sec:dune}, we give an overview over the experiment. Section~\ref{sec:sn} will discuss the expected neutrino signal from a supernova burst while section~\ref{sec:time} focusses on the time structure of that signal.

\section{The Deep Underground Neutrino Experiment (DUNE)}\label{sec:dune}

LAr TPCs are a recent neutrino detection technology which offers precision 3D imaging with mm-scale resolution, excellent energy measurement through active calorimetry and accurate particle ID via energy loss (dE/dx) and event topology.
DUNE aims to scale this technology up to the 10\,kton scale which would allow using it as the far detector in a long-baseline experiment and give much higher statistics for other areas, making it highly competitive with detectors based on other technologies.

DUNE is a modular experiment consisting of four independent detectors that will be located at the 4850\,ft level of Sanford Underground Research Facility (SURF), corresponding to 4300 meters water equivalent.

Two detector designs are currently under development: a single phase reference design, similar to previous LAr TPC detectors, and a dual phase design that includes a thin layer of gaseous argon at the top of the detector where ionization electrons will get amplified and collected, resulting in a higher signal-to-noise ratio.
Both detector designs would be located in a cryostat with internal dimensions $15.1(w)\times 14(h)\times 62(l)\,\text{m}^3$ and have a fiducial mass of 10\,kton, resulting in 40\,kton fiducial mass for the whole experiment.

Work to prepare excavation of the caverns is currently ongoing; the first module is planned to start data-taking in 2024 with the other modules following soon thereafter.

\section{Supernova Observation with DUNE}\label{sec:sn}

\begin{figure}[htb]
\centering
\includegraphics[width=24pc]{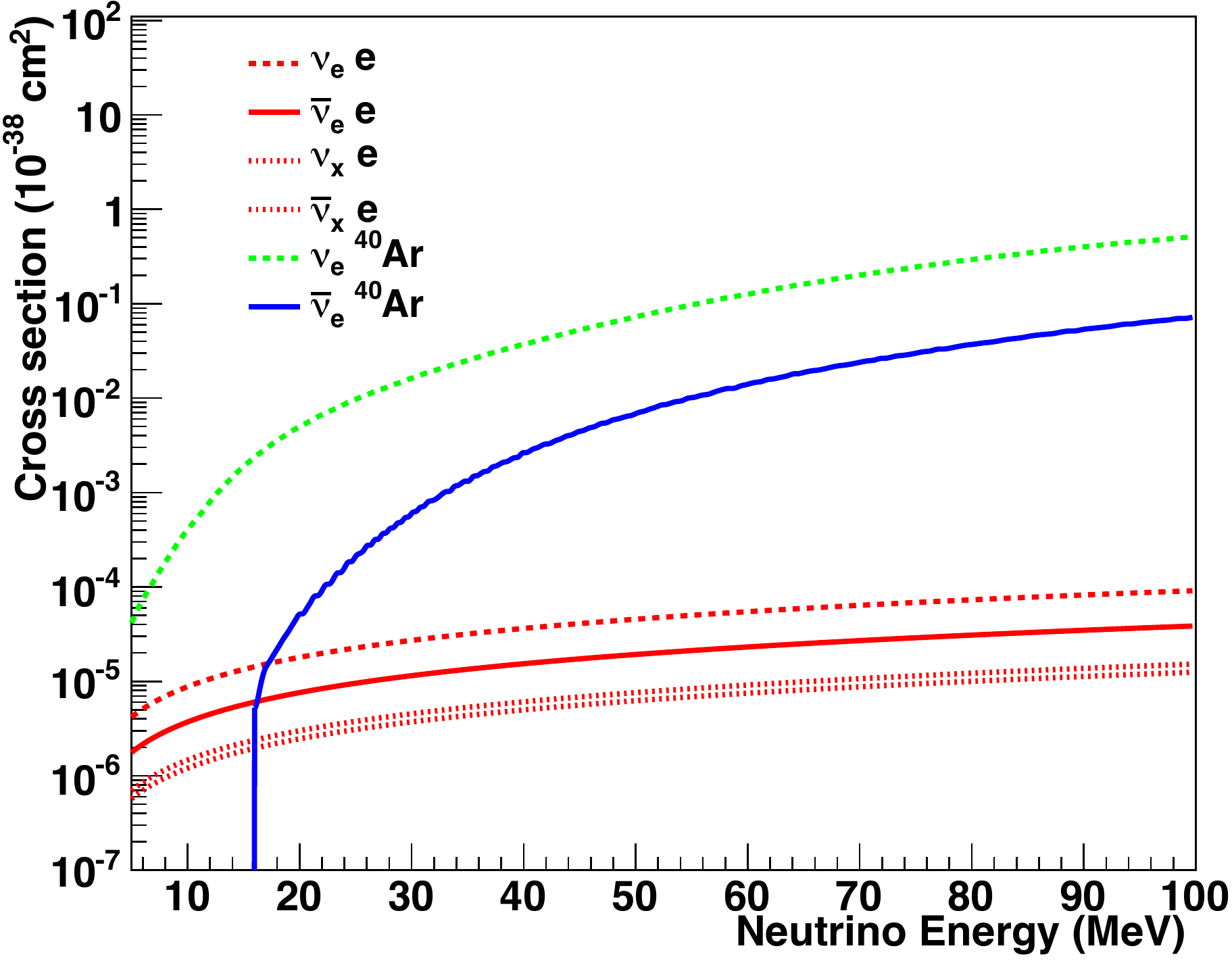}
\caption{Cross sections for three main interaction channels of supernova neutrinos in argon: $\nu_e + \/^{40}\text{Ar} \rightarrow e^- + \/^{40}\text{K}^*$ ($\nu_e$CC, dashed green), $\bar{\nu}_e + \/^{40}\text{Ar} \rightarrow e^+ + \/^{40}\text{Cl}^*$ ($\bar{\nu}_e$CC, solid blue) and $\nu + e^- \rightarrow \nu + e^-$ (ES, various red)~\cite{GilBotella:2003sz}.}
\label{fig:argon_xscn}
\end{figure}

In the energy range of 5--50\,MeV typical for supernova neutrinos, there are three main neutrino interaction channels in liquid argon  (see figure~\ref{fig:argon_xscn}), with the $\nu_e$CC channel containing 80--90\,\% of events. This makes DUNE particularly sensitive to the $\nu_e$ component of the supernova neutrino flux and thus highly complementary to other detector technologies which are most sensitive to the $\bar{\nu}_e$ component.

\begin{figure}[htb]
\centering
\includegraphics[width=32pc]{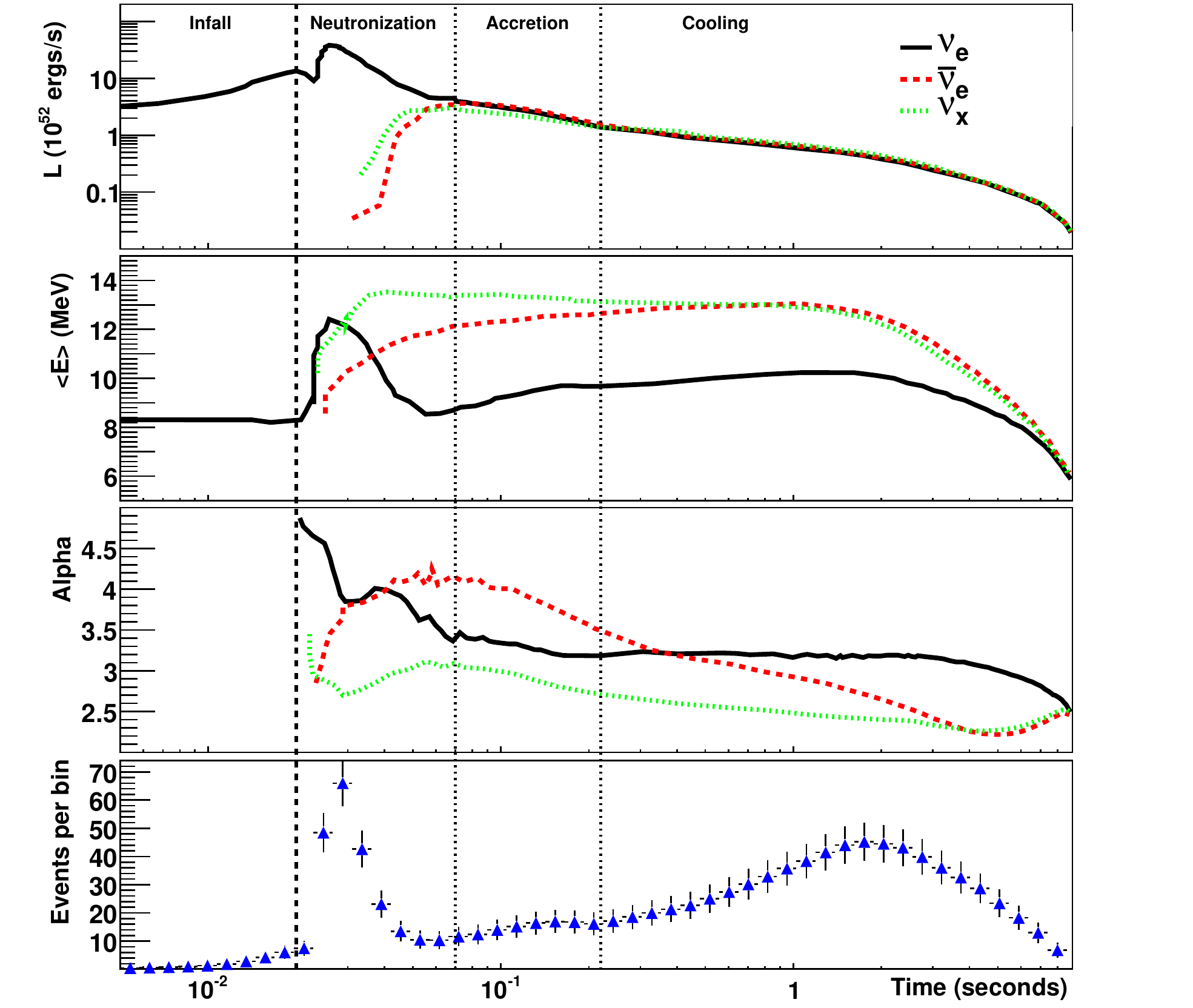}
\caption{Neutrino signal from an electron-capture supernova~\cite{Huedepohl:2009wh} at 10\,kpc, assuming no oscillations. The first three plots show the luminosity, average neutrino energy and the $\alpha$ (pinching) parameter as a function of time. The fourth (bottom) plot shows the total number of events (mostly $\nu_e$) expected in 40\,kt of liquid argon, calculated using SNOwGLoBES~\cite{snowglobes}. Note the logarithmic binning in time; error bars are statistical.}
\label{fig:garching}
\end{figure}

Overall, for a galactic supernova at a fiducial distance of 10\,kpc, DUNE is expected to observe $\sim$3500 events within $\sim$10\,s. Figure~\ref{fig:garching} shows the signal from one simulation.
For a more distant, SN1987a-like supernova in the Large Magellanic Cloud, we expect $\sim$50 events in DUNE, while a supernova in the Andromeda galaxy is expected to produce $\sim$1 event and would thus likely not be detectable.

The total neutrino flux from all remote supernovae in the history of the universe is known as the diffuse supernova neutrino background (DSNB). Depending on the theoretical model, DUNE might see a few DSNB events per year in the range between 16--40 MeV. Whether these could be detected above backgrounds is still unclear and will require a detailed study in the future.

\section{Time Structure of the Supernova Burst Signal}\label{sec:time}

Various mechanisms that could produce time- and energy-dependent features in the neutrino signal have been proposed.

This includes hydrodynamic features of the shock wave, like the standing accretion shock instability (SASI), as well as particle physics effects like self-induced flavor transitions due to the high neutrino density inside the supernova. Unfortunately, these effects are often highly model-dependent and while they are an area of intense theoretical study, no robust experimental signatures have yet been identified~\cite{Scholberg:2017czd}.
After the next galactic supernova, looking for evidence of these effects (and, potentially, dis\-entangling them) in the neutrino signal will be a high priority.

\begin{figure}[htb]
\centering
\includegraphics[width=36pc]{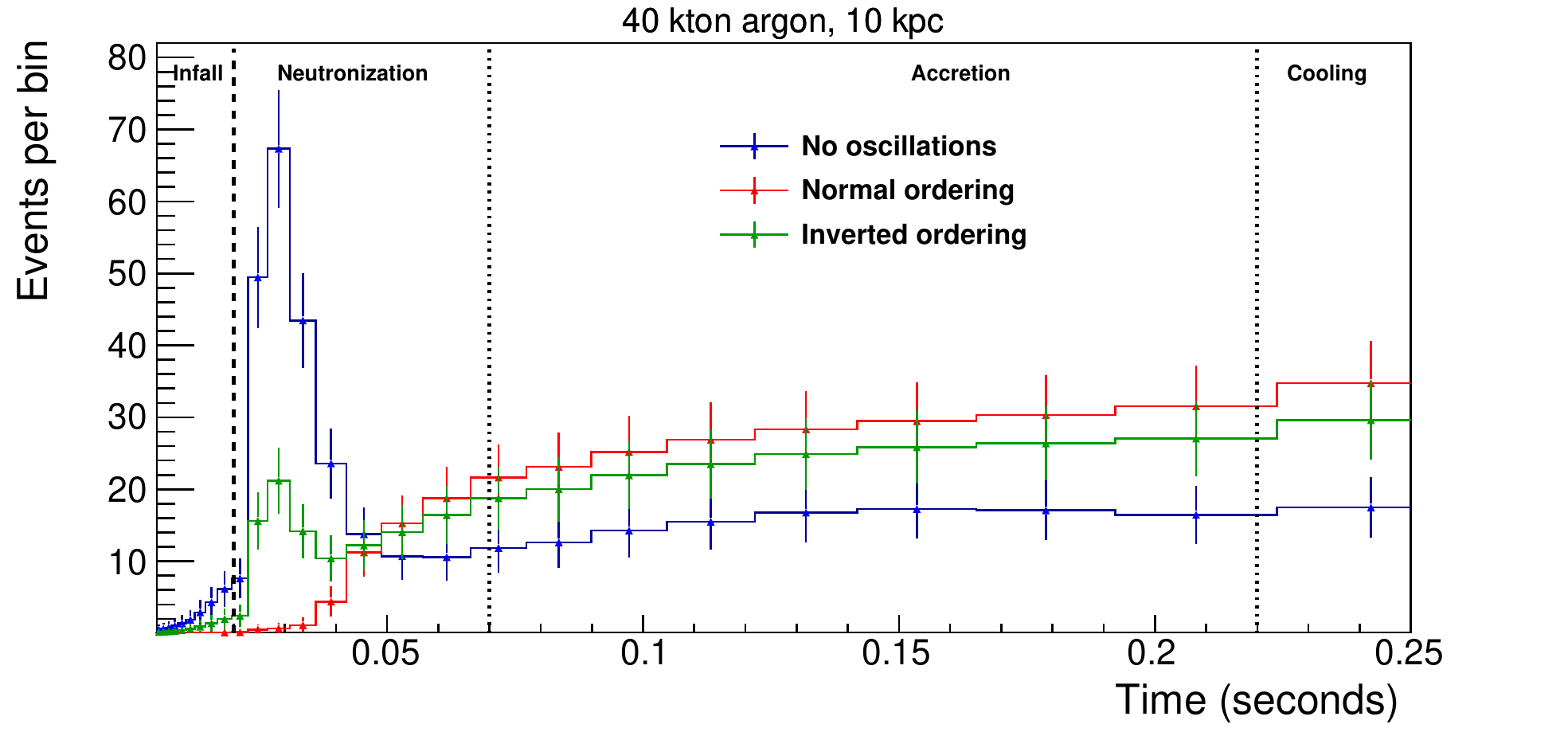}
\caption{Expected time-dependent signal for a specific flux model for an electron-capture supernova~\cite{Huedepohl:2009wh} at 10\,kpc in 40\,kton of argon, for times up to the start of the cooling phase. The expected signal assuming MSW oscillations only (i.e. no collective effects) is shown under the assumptions of normal and inverted hierarchy. The total number of events per bin (dominated by electron neutrino flavor) is shown in logarithmic bins. Plot made with SNOwGLoBES.}
\label{fig:early_time_argon}
\end{figure}

On the other hand, the neutronization burst at $\sim$10\,ms after core bounce is a nearly model-independent feature consisting solely of $\nu_e$ that is present with a comparable duration, shape and intensity across a wide range of computer simulations.
Neutrinos in the neutronization burst undergo MSW oscillations on their way through the outer layers of the supernova, making the observed signal dependent solely on the mass ordering as shown in figure~\ref{fig:early_time_argon}.


\end{document}